\documentclass{ws-ijmpe}

\usepackage{amssymb}
\usepackage{graphicx}

\begin{document}

\markboth{G.Papp, P. L\'evai, G. Fai}{Intrinsic transverse momentum in
di-hadron correlations}

\catchline{}{}{}{}{}

\title{\bf INTRINSIC TRANSVERSE MOMENTUM IN DI-HADRON CORRELATIONS}

\author{G\'ABOR PAPP}

\address{Institute for Physics, E\"otv\"os University, P\'azm\'any P. s. 1/A, Budapest, H-1117, Hungary\\
pg@ludens.elte.hu}

\author{P\'ETER L\'EVAI}

\address{RMKI, 
P.O. Box 49, Budapest, H-1525, Hungary\\
plevai@rmki.kfki.hu}

\author{GEORGE FAI}

\address{CNR, 
Department of Physics, Kent State University, Kent, OH-44242, USA\\
fai@cnr8.kent.edu}

\maketitle

\begin{history}
\received{(received date)}
\revised{(revised date)}
\end{history}

\begin{abstract}
We study di-hadron correlations in proton-proton collisions at
$\sqrt{s} = 200$ GeV and interpret experimental data in terms of a
fragmentation width and a momentum imbalance. A fragmentation width
of $580 \pm 50$ GeV/c is obtained, and the
measured momentum imbalance gives
an `intrinsic' transverse momentum width of partons in the proton of
$2.6 \pm 0.2$ GeV/c.
Consequences to heavy ion collisions are discussed.
\end{abstract}

\section{Introduction}

Recent high-statistics runs at RHIC allow the study of two-particle
correlations, with pronounced suppression pattern for Au+Au
collisions, augmenting information from jet quenching, and are a
valuable tool of jet tomography. Questions raised by di-hadron
studies motivated programs of three-body correlation measurements to
clarify the collective dynamics of nuclear collisions.

To argue about any suppression  of the correlation data in heavy-ion
collisions a proton-proton ($pp$) reference is needed, preferably at
the same energy. Correlation data in $pp$ at $\sqrt{s}=200$ GeV have
become available recently~\cite{unknown:2006sc}. Our goal here is to
provide a physical picture of this rich data set on near and away
side correlations. To interpret the data, we want to stay as close
as possible to Ref.~\cite{unknown:2006sc} with the ingredients of
our calculation and since fragmentation of pions are best
established so far phenomenologically, we focus on pion
correlations.

\section{Model}
\label{sec_mod}
The single-pion inclusive production cross section
can be written as
\begin{equation}
\frac{d\sigma_{\pi}}{p_T dp_T} = \int \frac{d\sigma_{j}}{p_T dp_T} D_j(z) dz
=\int \frac{d\sigma_{j}}{{\hat p}_T d{\hat p}_T} D_j(z)\frac{dz}{z^2}
=\int f_j D_j(z)\frac{dz}{z^2} \ ,
\end{equation}
where $d\sigma_{j}/{\hat p}_T d{\hat p}_T$ refers to the differential jet
cross section in terms of the parton transverse momentum ${\hat p}_T$,
and $z$ is the momentum fraction carried by the observed hadron.
The quantity $f_j$ is a parton (jet) distribution averaged over quarks,
antiquarks, and gluons, and  $D_j(z)$ is an average fragmentation
function. Using $f_j ({\hat p}_T) \propto {\hat p}_T^{-n}$ and
$D_j(z) \propto z^{-\alpha} (1-z)^\beta (1+z)^{-\gamma}$ with parameters
$n=7.4$, $\alpha=0.32$, $\beta=0.72$, $\gamma=10.65$\cite{unknown:2006sc},
we reproduce the measured pion spectra~\cite{Adler:2003pb} for $p_T > 3$ GeV,
which is satisfactory for our present analysis.

On this basis, we constructed a simple model to describe two-particle
correlations within a jet and between back-to-back jets starting from a
hard $2 \rightarrow 2$ parton-parton collision. For
two pions produced from the same parton (near side correlation)
the two-particle cross section can be written as
\begin{equation}
d\sigma_{\pi_1 \pi_2}^{near}
 = \int_0^1 dz_1 \int_0^1 dz_2
 \ d\sigma_j \ D_{j1}(z_1) D_{j2}(z_2) \ \Theta (1-z_1-z_2) \ ,
\label{2dsig0}
\end{equation}
while for the away-side (two pions produced from {\it different}
partons)
\begin{equation}
d\sigma_{\pi_1 \pi_2}^{away} = \int_0^1 dz_1 \int_0^1 dz_2
 \ d\sigma_j \ D_{j1}(z_1) D_{j2}(z_2) \ ,
\label{2dsig1}
\end{equation}
where the momentum fractions $z_1$ and $z_2$ describe the relation
between the parton and hadron momenta, $p^*_{T1} = z_1 {\hat p}_T$
and $p^*_{T2} = z_2 {\hat p}_T$. Moreover, due to fragmentation, the
produced hadrons acquire a random transverse momentum component. In
addition, the momentum imbalance ($K_T$) between the produced
partons due to intrinsic transverse momentum, gluon radiation, or
any other $2 \rightarrow 3$ process will generate a more complicated
kinematic situation, where, in the case of back-to-back jets, ${\hat
p}_{T1}$ and ${\hat p}_{T2}$ are already not collinear.

Considering hadron 1 as the trigger hadron and hadron 2 as the associated hadron,
after kinematic transformations,
\begin{eqnarray}
\frac{d\sigma_{\pi_t \pi_a}}{d p_{Tt} d{\Delta \phi} d p_{Ta} }
&=& J^* \frac{d\sigma_{\pi_t \pi_a}}{d {p}^*_{Tt} dz_t d{p}^*_{Ta} }
= J \cdot f_j D_{jt}(z_t) D_{ja}(z_a) \ \Theta(1-z_t-z_a) \, ,
\label{Xdphi}
\end{eqnarray}
where $\Delta \phi$ is the azimuthal angle difference between the
trigger and associated hadrons, and $J$ and $J^*$ represent the
proper Jacobi determinants of the variable transformations (without
the theta function for away-side).

We have developed a Monte-Carlo based calculation to model
Eq.~(\ref{Xdphi}). We use Gaussian distributions and uniformly
distributed random angles for the fragmentation transverse momenta
and the momentum imbalance $K_T$. The widths of these distributions
that best fit the data\cite{unknown:2006sc} are extracted.

\section{Results}
\label{sec_res}

\begin{figure}[h]
\centerline{\includegraphics[width=0.82\textwidth,
viewport=0 16 370 370]{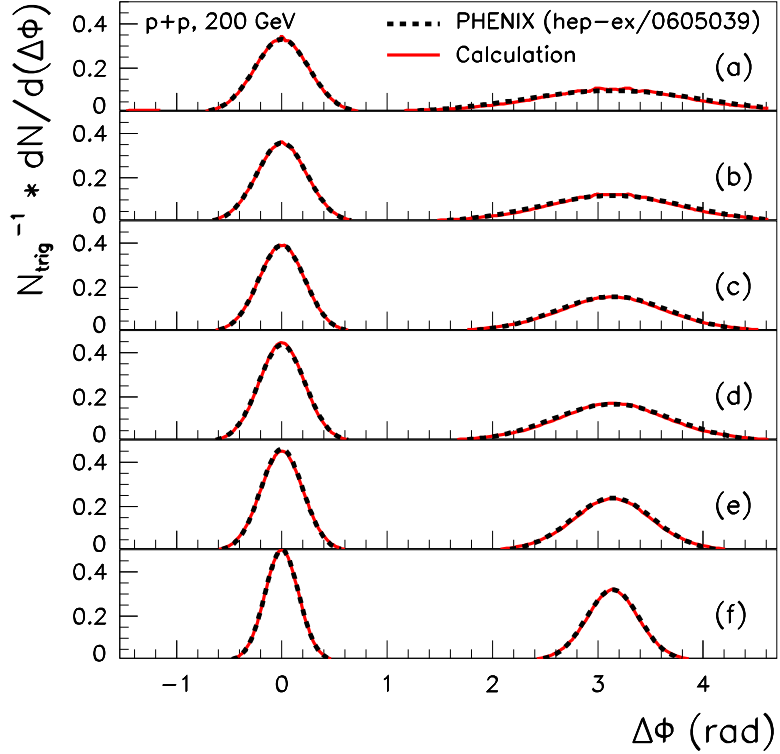} }
\caption{Di-hadron correlations in $pp$
collisions at $\sqrt{s} = 200$~GeV
based on the width of the fragmentation and $K_T$ distributions
(full lines) compared to data from Ref.~\protect\cite{unknown:2006sc}
(dashed lines). }
\label{fig1}
\end{figure}
We have calculated di-hadron correlation functions in $pp$
collisions at $\sqrt{s} = 200$ GeV at various trigger and associated
transverse momenta and compared the results to available
experimental data\cite{unknown:2006sc} to extract the widths of the
fragmentation transverse momentum distribution and the $K_T$
distribution. A sample comparison is displayed in Fig.~1, where the
dashed lines indicate a fit through the data points, and the full
lines represent our calculations. The associated transverse momentum
is kept in the
$1.4 \leq p_{Ta} \leq 5.0$ GeV/c range, 
while the trigger transverse momentum is increasing from $2.5 \leq
p_{Tt} \leq 3.0$ GeV/c to $6.5 \leq p_{Tt} \leq 8.0$ GeV/c moving
down through the panels. The ``pedestal'' of the correlation
functions has been cut off. The agreement is very good, although
small deviations are visible in a more magnified view.

Figure 2 shows the values of the widths of the fragmentation and
`intrinsic' momentum distributions in given $p_{Tt}$ windows as
functions of $p_{Ta}$. The fragmentation width appears to be
constant, $\sqrt{\langle j_T^2 \rangle} = 580 \pm 50$ GeV/c,
independent of $p_{Ta}$ and $p_{Tt}$. The $k_T$ width shows a
dependence on $p_{Tt}$ (similarly to a simpler
treatment~\cite{Levai:2005fa}), displayed for two limiting data sets
in Fig.~3.

\begin{figure}[h]
\centering
\includegraphics[width=0.48\textwidth,height=5truecm,viewport=0 170 370 374]{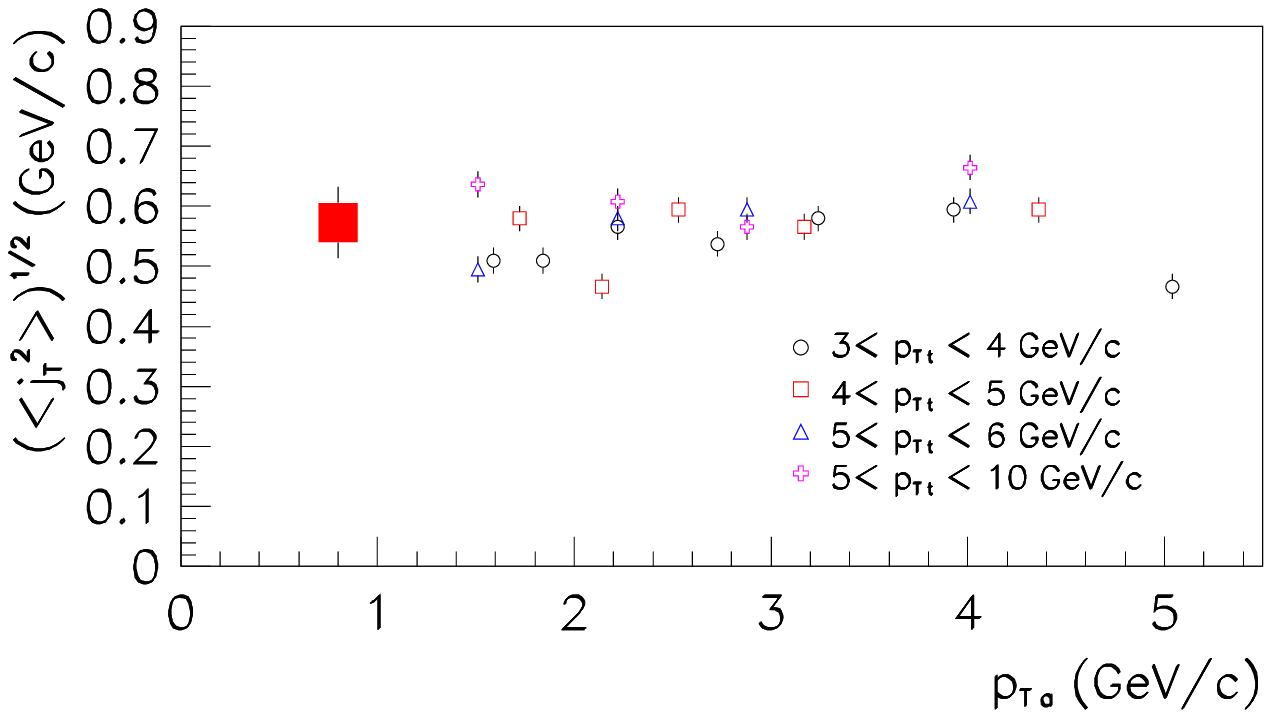}
\label{fig2} \hfill
\includegraphics[width=0.48\textwidth,height=5truecm,viewport=0 170 370 374]{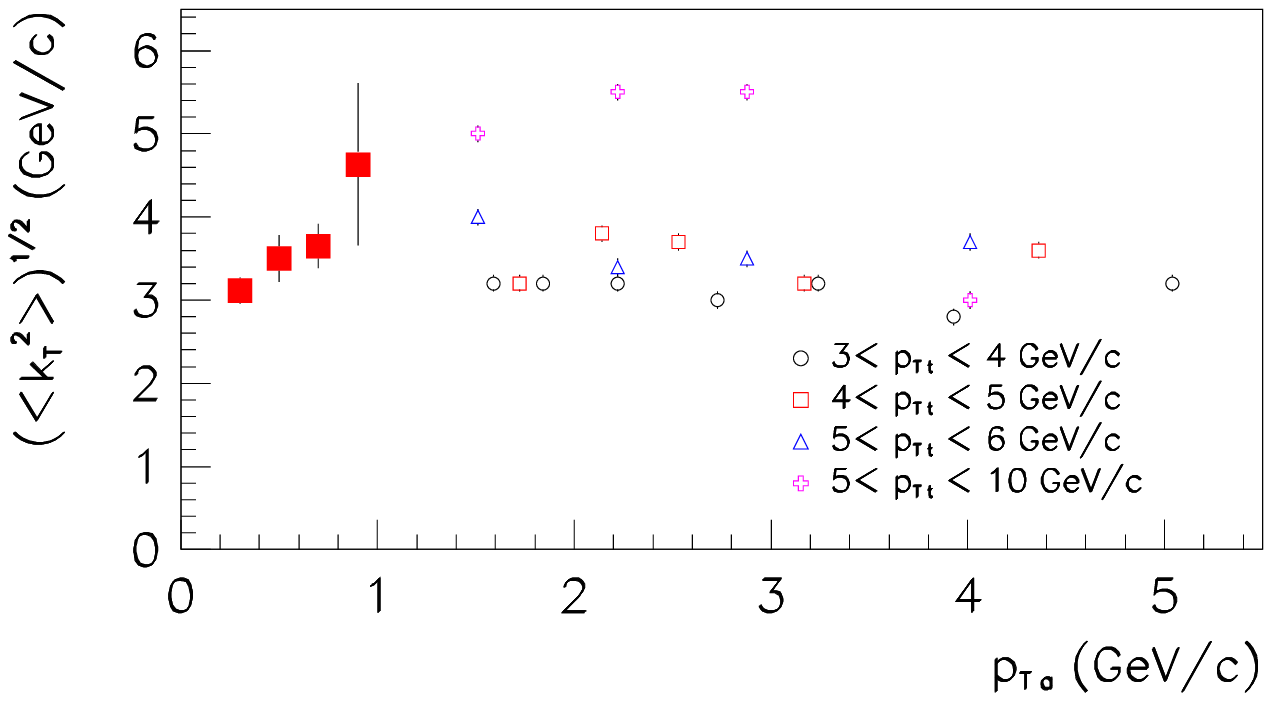}
\label{fig3} \caption{Best fit values of the Gaussian fragmentation
width (left) and $k_T$ width (right) to reproduce the near and away
side peaks in given $p_{Tt}$ windows as a function of $p_{Ta}$.
Averages for $p_{Tt}$ windows (only the grand total average on the
left) are indicated by the large filled squares. Data are from
Ref.~\protect\cite{unknown:2006sc}. }
\end{figure}

\begin{figure}[h]
\centering
\includegraphics[width=0.65\textwidth,viewport=0 16 370 370]{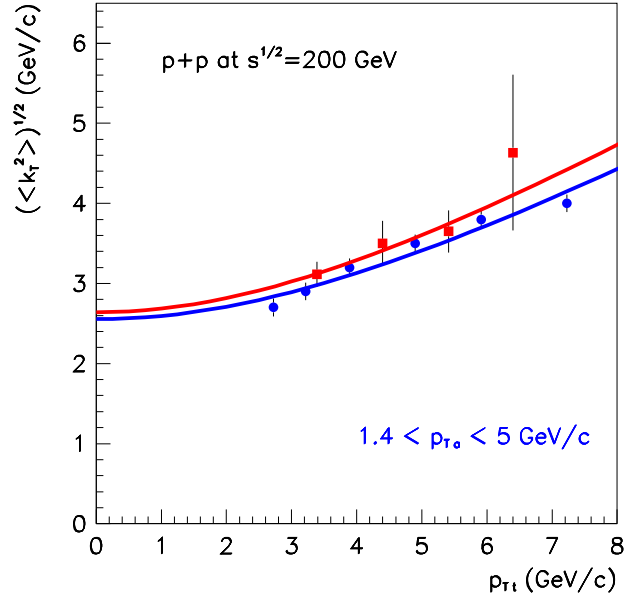}
\caption{Width of the `intrinsic' transverse momentum distribution
as a function of the trigger transverse momentum for two sets of
data. }
\label{fig4}
\end{figure}
%
%
The spread between the two least-square fitted curves in Fig.~3 is
indicative of the uncertainty in our procedure.
The behavior of the `intrinsic' transverse momentum width as a function of
$p_{Tt}$ can be understood in terms of its composition\cite{Boer:2003tx}. In addition
to the `true' intrinsic transverse momentum of partons in the proton, there is a
component from soft gluon radiation that can be handled via resummation, and a
higher-order contribution which is expected to grow with trigger transverse momentum:

\begin{equation}
\frac{\langle p_T^2 \rangle_{pair}}{2} = \langle k_T^2 \rangle =
\langle k_T^2 \rangle_{intrinsic} + \langle k_T^2 \rangle_{soft} +
\langle k_T^2 \rangle_{higher-order} \,\, .
\label{kTcomp}
\end{equation}
A measure of the `intrinsic' transverse momentum of partons in the proton (in which
we include the soft gluon radiation component) can be read from Fig.~\ref{fig4}
extrapolating to $p_{Tt} = 0$. This leads to
$\sqrt{\langle k_T^2 \rangle} = 2.6 \pm 0.2$ GeV/c, in agreement with the value
arrived at in Ref.~\cite{unknown:2006sc}, albeit by a 
different argument.

This value for the 'intrinsic' transverse momentum is somewhat
larger than the value obtained from our earlier analysis of
one-particle pion spectra~\cite{Yi02,Levai07} and other
analyses~\cite{Ziel98,BGG07}. However, it fully supports the
inclusion of such a quantity in the study of one-particle hadron
spectra and hadron-hadron correlation data. Without this momentum
imbalance no coherent picture of high-$p_T$ particle production in
proton-proton collisions can be established, neither in leading
order, nor in next-to-leading order perturbative QCD calculations.

Furthermore, this momentum imbalance exists in hard particle
production in proton-nucleus and heavy ion collisions, also. The
measured double-hump structure of away side hadron-hadron
correlation data in Au-Au collisions~\cite{STARaway,PHENIXaway} has
been explained by the formation of a shock wave in the highly
excited matter ("Mach-cone")~\cite{Stocker,Ruppert,Casald}. The
presence of the 'intrinsic' transverse momentum immediately after
the formation of the jet pairs contributes to the widening of the
away-side correlation peak in such a way that the contribution from
shock-wave formation becomes smaller and the qualitative description
will be different (i.e. the opening angle of the Mach-cone and the
obtained speed of sound for the excited matter). The forthcoming
quantitative analysis of heavy ion experiments requires more precise
understanding of the away-side data in proton-proton collisions.

\section{Conclusion}
\label{sec_conc}

We have developed a constructive method for taking into account the fragmentation width
and the momentum imbalance in the treatment of di-hadron correlations. The model has
the flexibility to treat any $K_T$ distribution. Our results have been tested against
recent PHENIX data on di-hadron correlations in $pp$ collisions. A value of
$2.6 \pm 0.2$ GeV/c is obtained for the width of the `intrinsic' transverse momentum
distribution. Future work includes generalization to $pA$ and $AA$ collisions,
inclusion of the pseudorapidity dimension of the correlations, and application in
jet quenching calculations.

\section*{Acknowledgements}

Supported in part by U.S. DE-FG02-86ER40251, U.S. NSF INT-0435701,
and Hungarian OTKA grants T043455, T047050, and NK062044.

\end{document}